\documentclass{eas}
\usepackage{amsmath,amssymb,epsfig,graphicx}
\usepackage{natbib}

\newcommand{\ben}{\begin{eqnarray}}
\newcommand{\een}{\end{eqnarray}}
\newcommand{\mean}[1]{\mbox{$\langle #1 \rangle$}}
\newcommand{\sigv}{\mean{\sigma v}}
\newcommand{\mchi}{m_{\chi}}
\newcommand{\pbar}{\mbox{$\overline{p}$}}
\newcommand{\pbars}{\mbox{$\overline{p}{\rm 's}$}}
\newcommand{\pos}{\mbox{$e^+$}}
\newcommand{\poss}{\mbox{$e^+{\rm 's}$}}

\newcommand{\Msun}{\mbox{$M_\odot$}}
\newcommand{\Rsun}{\mbox{$R_\odot$}}
\newcommand{\xsun}{\mbox{$\vec{x}_\odot$}}
\newcommand{\rhosun}{\mbox{$\rho_\odot$}}
\newcommand{\Mcl}{\mbox{$M_{\rm cl}$}}

\newcommand{\phicltot}{\mbox{$\phi_{\rm cl}^{\rm tot}$}}
\newcommand{\cvir}{\mbox{$c_{\rm vir}$}}
\newcommand{\boost}{\mbox{${\cal B}$}}
\newcommand{\vconv}{\mbox{$V_{\rm c}$}}
\newcommand{\gprop}[1]{\mbox{${\cal G}( #1 )$}}
\newcommand{\gtilde}[1]{\mbox{$\tilde{\cal G}( #1 )$}}

\newcommand{\lcdm}{$\Lambda$-CDM}

\begin{document}

\title{On the antimatter signatures of the cosmological dark matter subhalos} 
\runningtitle{On the antimatter signature of the cosmological dark matter 
  subhalos}

\author{Julien Lavalle}
\address{Dipartimento di Fisica Teorica, Universit\`a di Torino \& INFN,
  Via Giuria 1, 10125 Torino -- Italia}
\email{lavalle@to.infn.it}

\begin{abstract}
  While the PAMELA collaboration has recently confirmed the cosmic ray 
  positron excess, it is interesting to review the effects of dark matter (DM)
  subhalos on the predicted antimatter signals. We recall that, according to 
  general subhalo properties as inferred from theoretical cosmology, and
  for DM with constant annihilation cross section, the enhancement cannot 
  be $\gtrsim$ 20 for the antimatter 
  yield. This bound is obviously different from that found for 
  $\gamma$-rays. We also recall some predictions for supersymmetric benchmark 
  models observable at the LHC and derived in the cosmological N-body 
  framework, showing in the meantime the existing discrepancy between 
  profiles derived from N-body experiments and the current observations of 
  the Milky Way.
\end{abstract}

\maketitle

\begin{figure*}[t]
\begin{center}
\includegraphics[width=0.3\columnwidth]{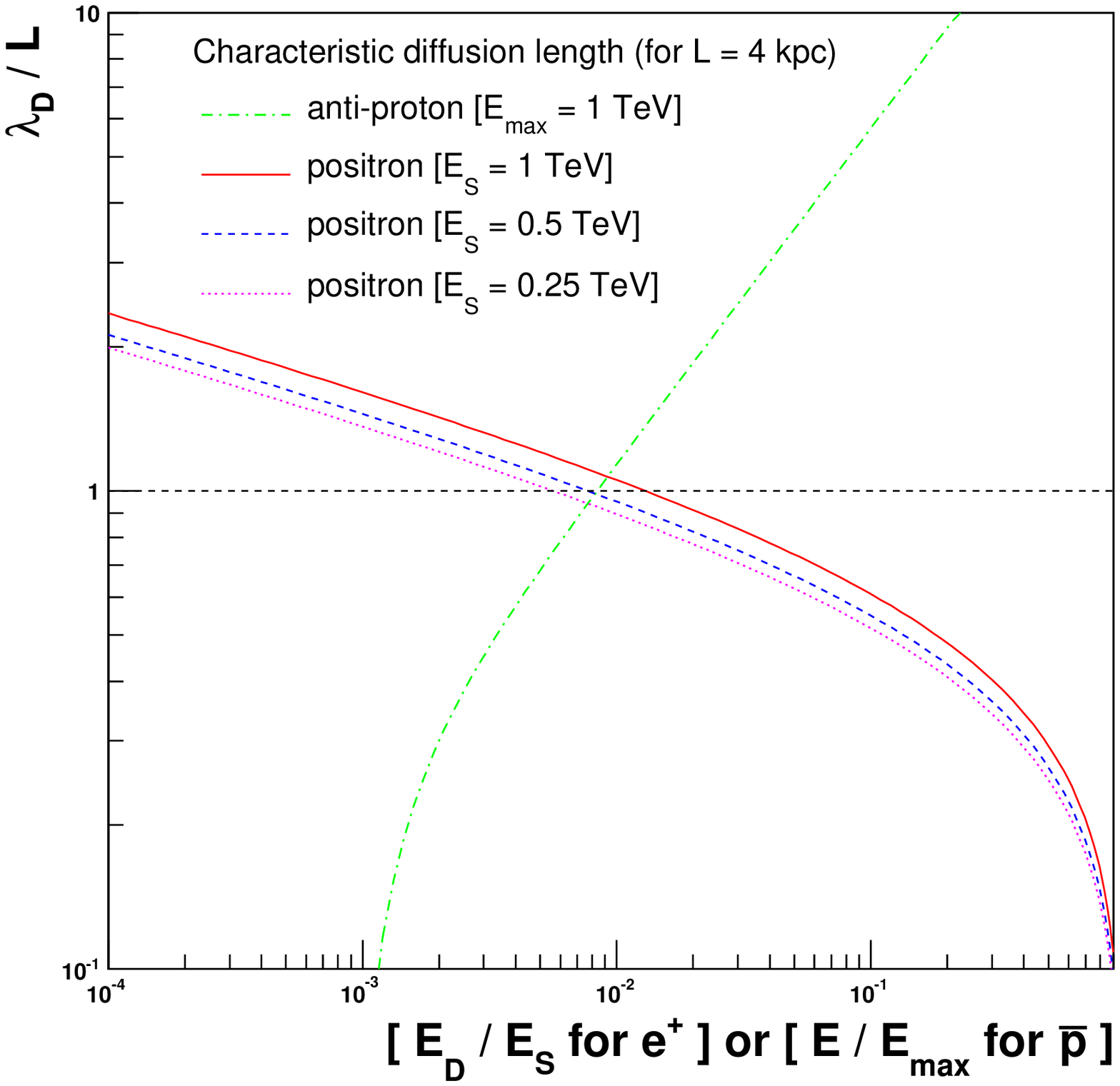}
\includegraphics[width=0.3\columnwidth]{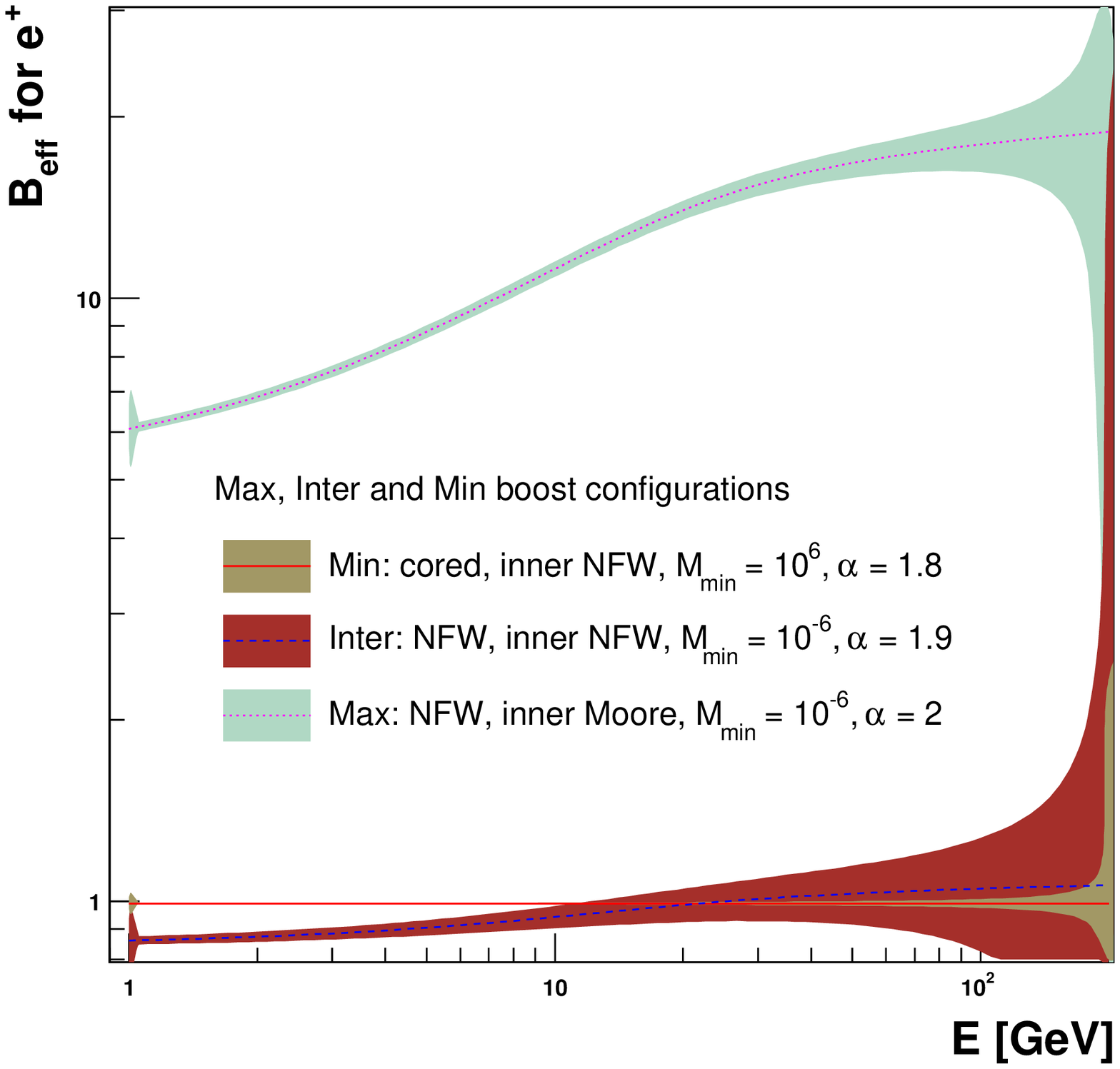}
\includegraphics[width=0.3\columnwidth]{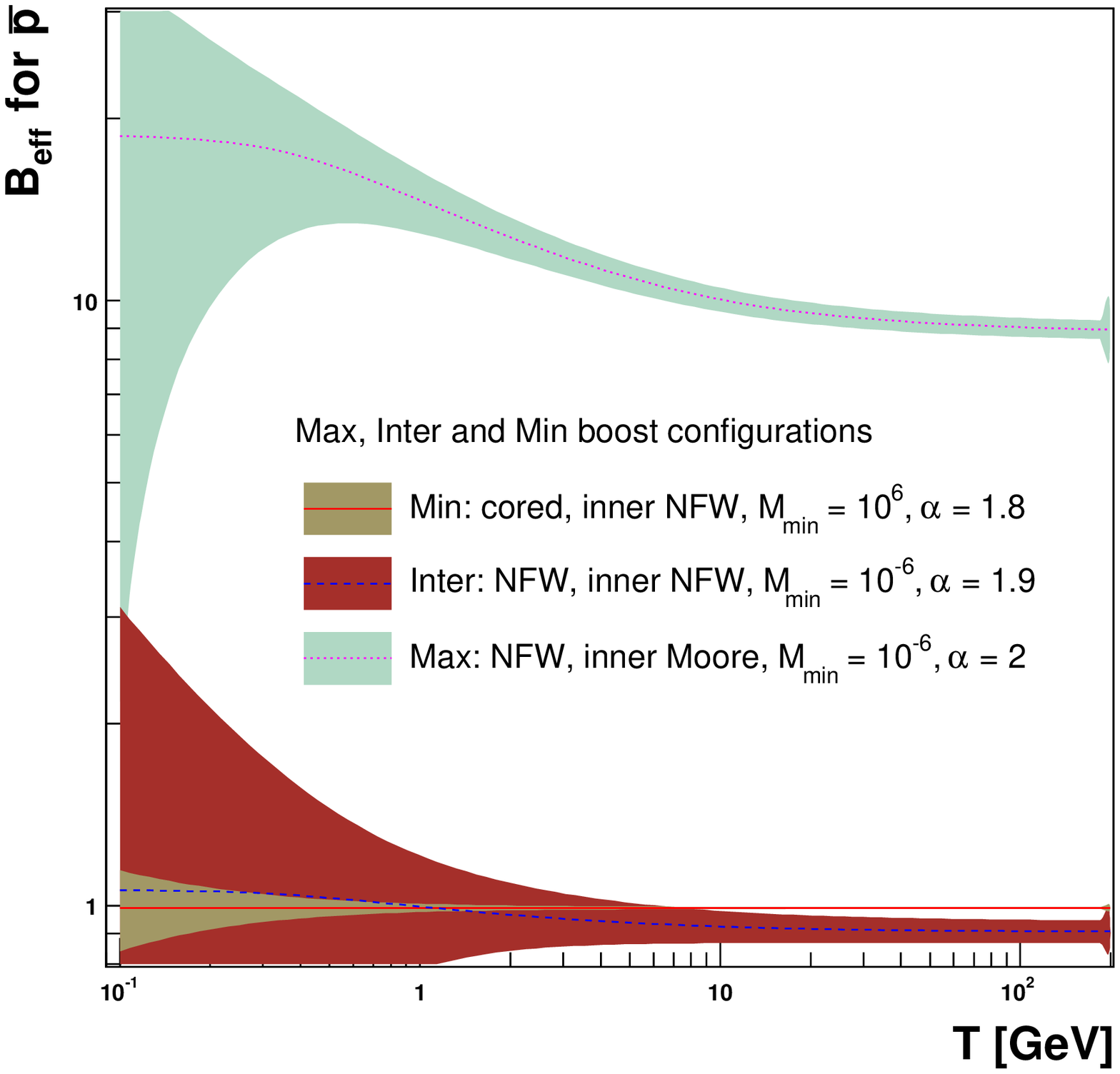}
\caption{\small Left: propagation scales for \pbars\ and \poss. The 
horizontal axis is detected energy $E_{\rm D}$ / injected energy $E_S$ for 
\poss, while only $E$ / 1 TeV for \pbars, since the latter do not lose energy. 
Middle: mean boost factors and associated variances obtained for 
\poss\ as functions of the energy at the Earth, for different subhalo 
models. The top curve corresponds to the \emph{maximal} model, which, as 
expected, is found to increase for decreasing propagation scales, but as the 
variance does. Right: the same for \pbars, where the energy dependence is 
reversed compared to \poss.} 
\label{fig:scales_and_boosts}
\end{center}
\vspace{-0.5cm}
\end{figure*}

\begin{figure*}[t]
\begin{center}
\includegraphics[width=0.3\columnwidth]{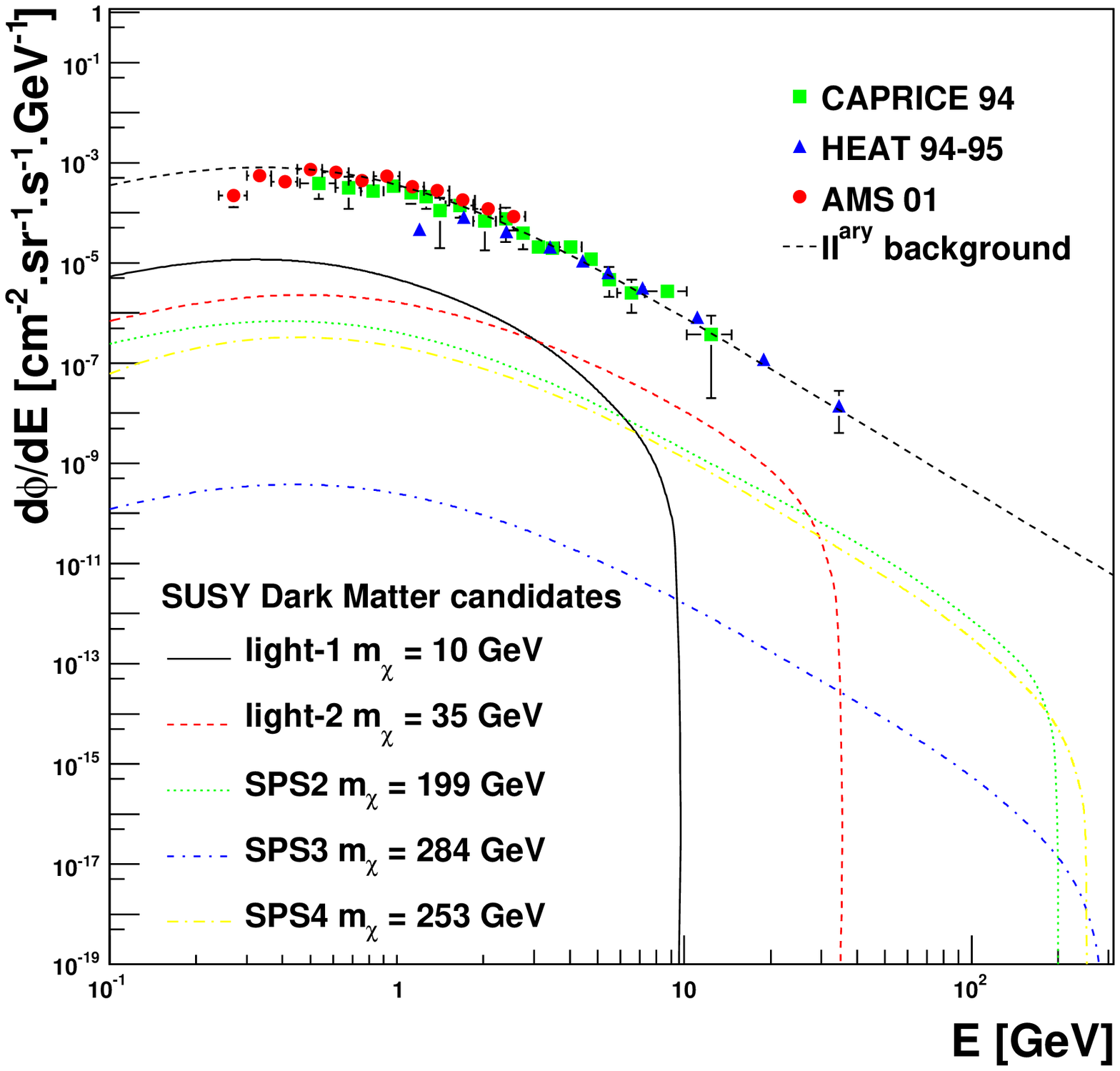}
\includegraphics[width=0.3\columnwidth]{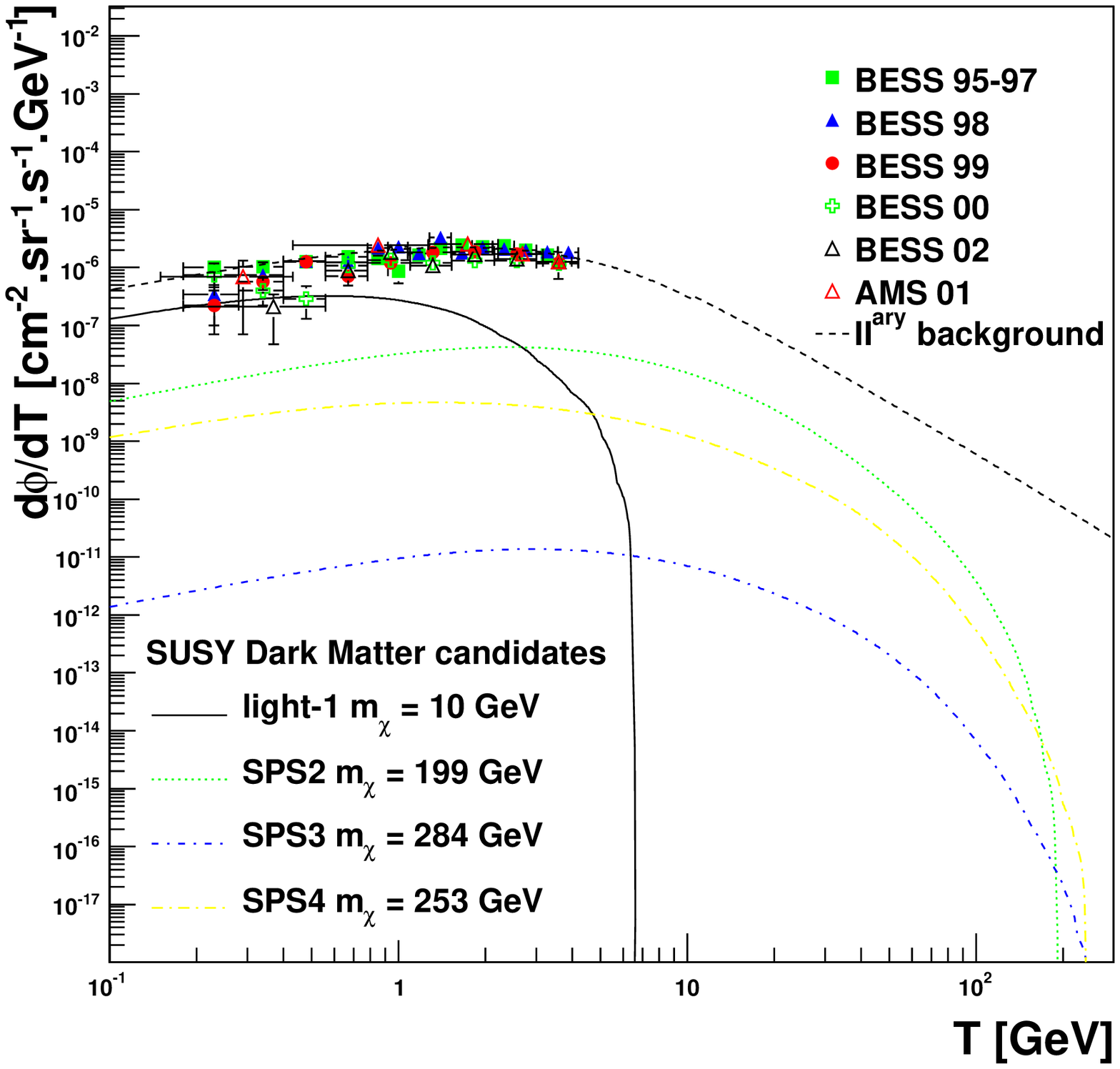}
\includegraphics[width=0.3\columnwidth]{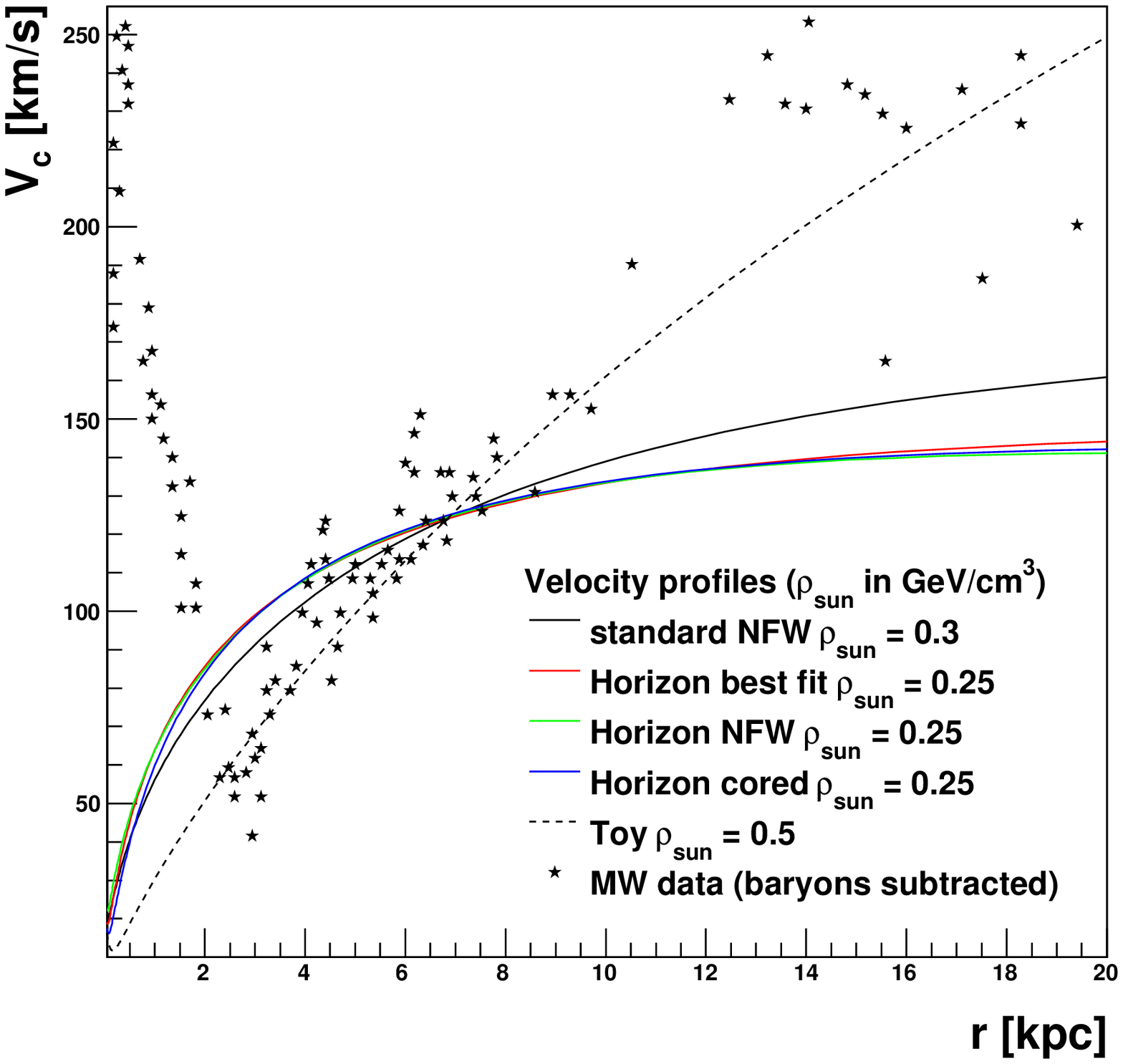}
\caption{\small Left: predictions for the \pos\ fluxes for some supersymmetric 
benchmark models as directly calculated from a N-body simulation. Middle: 
same for \pbars. Right: baryon-subtracted radial velocity curves for the 
Galaxy, where are also reported the DM contributions from different density 
profiles as found in cosmological N-body simulations.}
\label{fig:susy_and_vcurves}
\end{center}
\vspace{-0.5cm}
\end{figure*}

The positron (\pos) cosmic ray (CR) fraction has recently drawn considerable 
attention due to the confirmation of a \pos\ excess reported by the PAMELA 
collaboration up to 270 GeV~\citep{2008arXiv0810.4995A}. While the amplitude 
of this excess is unclear according to the most recent predictions of the 
secondary \pos\ flux at the Earth~\citep[where we have emphasized the role 
of the $e^-$'s]{2008arXiv0809.5268D}, dark matter (DM) annihilation has been 
again proposed as a source of such \poss, as it was already the 
case with the less clear HEAT excess ten years ago~\citep{1997ApJ...482L.191B,
1999PhRvD..59b3511B}. However, this primary contribution usually needs to 
be amplified to fit the data, and most of the authors have invoked 
some ad hoc boost factor coming from the presence of DM substructures in 
the Milky Way (MW), which should indeed increase the average annihilation rate.
Such subhalos are expected in the frame of hierarchical structure formation, 
and well resolved down to $\sim 10^5\Msun$ in cosmological N-body simulations 
as performed in the \lcdm~scheme. Here we shortly review the effects of DM 
clumps on the antimatter yield, focusing on \poss\ and antiprotons (\pbars). 
We will not discuss the PAMELA \pos\ fraction, since it is worth waiting for 
the release of the $e^{\pm}$ spectra themselves. The results and figures that 
we present here have been derived 
in~\citep{2007A&A...462..827L,2008A&A...479..427L,
2008PhRvD..78j3526L}, to which we refer the reader for more details and 
references.
{\bf Relevant scales for \poss\ and \pbars}.
As the DM distribution is usually found to exhibit a steep 
spatial dependence in N-body simulations, the knowledge of the typical CR 
propagation scale is useful: it characterizes the spatial 
extension over which the observed signal at the Earth is integrated, and
depends on both the propagation model and the CR species. 
In the GeV-TeV energy range, CRs diffuse on the 
Galactic magnetic turbulences, which ensure their confinement in the 
MW in a region which can be extended by a few kpcs above and below 
the Galactic disk. They interact through various processes with the 
interstellar medium (ISM) and/or the interstellar radiation field (ISRF). When 
propagating, \pbars\ mostly experience spallation in the 
disk, where the ISM is located, and convection outward the disk. 
These processes are efficient at low energies $\lesssim 5$ GeV, 
while energy losses are almost irrelevant; reacceleration can 
also be neglected. \poss\ obey to the same spatial diffusion as \pbars, but 
lose very quickly their energy, mostly through scattering on the 
ISRF. Their propagation is thus mainly set by the synchrotron and inverse 
Compton energy losses, of typical timescale $\tau\sim 300$ Myr, which occur in 
the whole diffusion zone. Other processes can be neglected without loss of 
accuracy above a few GeV. This allows to infer the characteristic propagation 
lengths for \poss\ and \pbars:
$\lambda_{\bar{p}} = \frac{K(E)}{\vconv}$ and
$ 
\lambda_{e^+}^2 = \frac{4 K_0\tau}{1-\delta}  
\left\{ \left(\frac{E}{E_0}\right)^{\delta-1} 
- \left( \frac{E_S}{E_0} \right)^{\delta -1} \right\}
$,
where $\vconv\sim 12$ km/s is the convection velocity, and $E_S$ is the 
\pos\ injected energy at the source, $K(E) = K_0(E/E_0)^\delta$ is the 
diffusion coefficient, with 
$\delta\sim 0.7$, $K_0\sim 1.12\times 10^{-2}{\rm kpc^2/Myr}$ being the 
normalization at $E_0 = 1$ GeV. The propagation length for a \pbar\ quickly 
increases with energy scaling like the diffusion coefficient 
$\propto E^{\delta}$. This is the opposite for \poss, since the propagation 
length increases only as a \pos\ loses its energy. This is illustrated in the 
left panel of Fig.~\ref{fig:scales_and_boosts}. For \poss, the 
characteristic propagation length is of order of a few kpcs, much lower than 
for \pbars, so the signal at the Earth comes from limited regions around, 
slightly enlarging as the detected energy decreases. Large integration 
volumes for \poss\ (\pbars) will be found at low (high) energy only, even 
reaching regions close to the Galactic center (GC). Therefore, local 
fluctuations of the injection rate mostly affect the high (low) energies 
for cosmic \poss\ (\pbars).
{\bf DM distribution in the MW and subhalos.}
The DM annihilation rate is set by the squared DM density. In the frame of 
\lcdm\ cosmology, the DM distribution is found to 
be rather scale invariant and close to universal in N-body simulations, when
baryons are not considered. This is the seminal result obtained 
by~\cite{1997ApJ...490..493N} (NFW), and confirmed, though with some 
scatter in the parameters, by many other groups since then. These scale 
invariant profiles are often given in the following spherical form:
  $\rho(r) = \rho_{0} ~\left(\frac{r}{r_0}\right)^{-\gamma} \left[
  \frac{1+\left(r_0/r_s\right)^{\alpha}}
  {1+\left(r/r_s\right)^{\alpha}}\right]^{\frac{\beta-\gamma}{\alpha}}$,
where $r_s$ is a scale radius, the index $0$ will refer here to the Sun 
location, such that $r_0=\Rsun=8$ kpc and 
$\rho_0=\rhosun = 0.3\;{\rm GeV.cm^{-3}}$. A NFW profile corresponds to 
$(\alpha,\beta,\gamma) = (1,3,1)$, and $r_s$ is usually found $\sim$ 20 kpc 
for MW-like objects. Although it makes sense to constrain the 
DM distribution from a top-down approach, it might sound risky to use 
such profiles, obtained without baryons, to make predictions at the 
sub-galactic scale, where baryons are expected to play a major role, 
but this has been done by many authors (including ourselves). Indeed, some 
observational constraints exist, so do many consistent dynamical modelings of 
the baryon component of the MW. In particular, \cite{2006CeMDA..94..369E} 
subtracted their baryon mass model to the velocity curves derived from CO and 
HI observations, and provided a set of data points on which one can directly 
constrain the DM contribution. In the right panel of 
Fig.~\ref{fig:susy_and_vcurves}, we show these data, on which we draw several 
density profiles. One is the NFW profile as employed for the MW, 
the others have been derived from fits on a N-body MW-like galaxy of the 
HORIZON project~\citep{2002A&A...385..337T}. These profiles clearly 
overestimate the central DM mass, so the DM annihilation rate close to the GC 
as well. While this will not affect that much the predictions for 
antimatter CRs, which are mostly set by the local density, this is 
instead expected to lower the $\gamma$-ray predictions. Now, we 
shortly turn to DM subhalos. For what concerns indirect detection of DM, the 
important features are their predicted spatial distribution, mass 
distribution and inner individual profile. There is no unique values predicted 
for these ingredients, so a way to estimate the theoretical uncertainties is to 
bracket each of them with the most extreme possibilities. As regards 
the spatial distribution, it is found antibiased with respect to the 
smooth profile in some N-body experiments~\citep{2007ApJ...657..262D}, which 
means $P_{\rm cl}(r)/\rho_{\rm s}(r) \propto r$, but we have also studied the 
case where subhalos track the smooth profile. The mass distribution
is usually found close to $M^{-\{ \alpha_M \simeq 2\}}$, as expected in the 
linear theory, and $\alpha_M\in[1.8,2]$ seems reasonable. The minimal mass is 
set by the free streaming scale of DM particles at the collapsing time. For 
weakly interacting particles with masses $\sim 100$ GeV, the minimal clump 
mass is $\sim 10^{-6}\Msun$~\citep{2006PhRvL..97c1301P}, which will be taken 
as a minimum here. The mass function can be assumed to be independent of the 
spatial distribution, so that the subhalo number distribution will be 
given by $dn/dM = N_{\rm tot}\cdot d{\cal P}_{\rm cl}(M)/dM \cdot 
d{\cal P}(r)/dV$, 
where $N_{\rm tot} = \mean{\Mcl} / \Mcl^{\rm tot}$ is the total number of 
clumps in the MW. The inner subhalo profile is mainly featured by its inner 
logarithmic slope $\gamma$ and the so-called concentration parameter. 
The latter is defined by $c_{\rm vir}\equiv r_{\rm vir}/r_{-2}$, i.e. the 
ratio of the subhalo virial radius to the radius at which the logarithmic 
slope of the profile is -2, and is found to decrease with the subhalo mass. 
Different models exist, usually fitted on N-body data, and we 
use two extreme configurations which encompass the wide range of 
possibilities: one optimistic extrapolated from~\cite{bullock_etal_01}, and 
a second \emph{minimal} from~\cite{eke_etal_01}, referred to as B01 and ENS01. 
Finally, the main quantity at stake when computing the antimatter flux is the 
intrinsic subhalo annihilation rate, which depends on its mass, 
concentration, and inner profile: $\xi(\Mcl,c_{\rm vir})\equiv
\int_{\rm cl}d^3\vec{x} (\rho_{\rm cl}/\rho_0)^2$, which has the dimension of a 
volume, and is normalized to the local rate. Note finally that clumps 
are not expected to be numerous in the central regions of the MW because of 
tidal disruption, and in any case not concentrated enough to dominate over the 
smooth annihilation rate there.
{\bf Fluxes and boost factors}.
The antimatter CR flux originating from the smooth DM halo 
is given by $\phi_{\rm s}(E) = {\cal S} \int d^3\vec{x} 
(\rho_{\rm s}(\vec{x})/\rho_0)^2 \cdot $ $ \int_{E} dE_S 
\gprop{\vec{x}_\odot,E \leftarrow \vec{x},E_S} \cdot dN/dE_S$, where 
${\cal S}\propto \sigv (\rho_0/\mchi)^2$ depends on the DM particle 
properties (whose annihilation cross section is assumed constant), 
$\gprop{f\leftarrow i}$ is the Green function associated with the 
CR propagation, and $dN/dE_S$ is the injected CR spectrum at 
the source. The flux due to subhalos is given by 
$\phicltot(E) = {\cal S} \sum_i \xi_i(\Mcl^i,\cvir) 
\gtilde{\vec{x}_\odot\leftarrow \vec{x}_i}\xrightarrow{\rm \small cont} 
{\cal S} \mean{N_{\rm cl}^{\rm tot}} \mean{\xi}\mean{\gtilde{\vec{x}_\odot}}$, 
where $\gtilde{f\leftarrow f}$ is the convolution of the Green function with 
the injected spectrum, and \mean{}\ means the average over the relevant 
subhalo distribution (space or mass). The boost factor is given by 
$\boost(E) = (1-f)^2 + \phicltot(E)/\phi_{\rm s}(E)$, where $f$ is the 
fraction of the local density in forms of clumps. It does depend on the 
energy, because of the energy dependence of the propagation 
scale, which sets the integration volume. It is independent from the 
injected spectrum only for \pbars. It is useful to derive the local 
asymptotic limit of the boost factor, since this allows the computation 
of its maximal value. It is valid at short propagation scales, i.e. at high 
(low) energy for \poss\ (\pbars):
$\boost_\odot \simeq 1 + \mean{N_{\rm cl}^{\rm tot}} \mean{\xi} 
d{\cal P}(\xsun)/dV$,
where $f$ is neglected. This expression does not depend on the 
CR species, and provides a good way to check numerical 
computations. We have calculated the boost factors and associated statistical 
variances for \pbars\ and \poss, by considering a flat spectrum for the former,
and a 200 GeV monochromatic line for the latter. We have studied different 
subhalo configurations, and different propagation models. The latter test 
mainly influences the variance, slightly the energy dependence of the central 
values, since it only affects the propagation scales. The former is instead 
responsible for a huge scatter in the single clump luminosity, but, 
interestingly, the central values still lie in a small range, since the smooth 
component is found to dominate the overall signal in most of cases. As shown 
in the middle and right panels 
of Fig.~\ref{fig:scales_and_boosts}, the maximal value obtained for both 
\poss\ and \pbars\ is $\sim 20$, which corresponds to the \emph{maximal} clump 
configuration: $\sim 10^{16}$ clumps in the MW, of minimal mass of 
$10^{-6}\Msun$, steep mass distribution $\propto \Mcl^{-2}$, spatially 
tracking the smooth NFW halo, with $r^{-3/2}$ inner profiles and B01 
concentrations. Such a configuration is theoretically very 
optimistic, so rather unlikely. Taking clumps of $10^{-9}\Msun$ would give an 
asymptotic value of $\boost_\odot\simeq 30$, which is still very weak, and 
considering a nearby very massive clump is unlikely and observationally 
constrained. The relative variance scales like $\sim\lambda_{\rm d}^{-3/2}$, as 
expected: the number of sources is greater in larger volumes. Moreover, at 
large $\lambda_{\rm d}$, the boost is close to 1 and the variance is small in 
any case, since the smooth contribution of the GC comes into play and 
dominates the signal, further shrinking the variance accordingly. We finally 
went beyond this analytical study by directly using a N-body galaxy from the 
HORIZON project, in which we could study 
more complex effects, e.g. density fluctuations (beside clumps themselves) or 
departure from spherical symmetry. We have quantified these effects to be 
small, and shown that a mere spherical smooth halo provides a good estimate of 
the antimatter fluxes. We show the predictions that we have derived for 
supersymmetric benchmark models observable at the LHC in 
Fig.~\ref{fig:susy_and_vcurves}.
{\bf Why is the boost for $\gamma$-rays different from that of CRs ?}
The argument is actually very simple. Indeed, for $\gamma$-rays, the signal is 
integrated along a line of sight, within a solid angle featured by the angular 
resolution of the telescope. It is therefore very natural to find that 
the boost for $\gamma$-rays is a function of the Galactic latitude: at low 
latitude, one looks toward the GC, where the smooth 
DM contribution dominates over that of subhalos, and the boost is 
very small; this situation is reversed at high latitude, where the 
boost is much more important, while the absolute flux remains 
small~\citep{1999PhRvD..59d3506B}. For antimatter CRs, one integrates 
the signal over very different regions which are much more concentrated 
around the Earth, where clumps are not expected to dominate. These regions are 
bounded by the CR propagation scale, which depends on both the species 
and the energy. It is therefore not surprising at all to find results 
different from $\gamma$-rays, even when adopting exactly the same subhalo 
configuration. For comparison, the maximal boost of 20 obtained for CRs 
would give instead a maximum of a few hundreds for $\gamma$-rays.
{\bf Conclusion}.
We have shown that subhalos are not expected to provide an important
enhancement to the antimatter flux expected from DM annihilation 
originating from a smooth description of the halo. Instead, one of the most 
important uncertainties affecting the predictions on the absolute fluxes is 
the local DM density. We have also stressed the origin of the 
differences between the boosts as computed for CRs and for 
$\gamma$-rays: the signals are integrated over quite different regions for 
those different messengers. Finally, we have reminded that the usual 
description of the Galactic DM halo, as very often employed for 
predictions of indirect astrophysical imprints of DM annihilation, is not 
consistent with the current observational constraints, and may lead to 
over-optimistic expectations. While the effect will be important 
for $\gamma$-rays, this has less impact on the predictions for antimatter 
CRs which are less sensitive to the yield from the central regions of 
the MW.
{\bf Acknowledgements}.
We warmly thank all of our collaborators having been involved in the 
derivation of the results summarized in this proceeding. In particular, we are
grateful to J. Pochon, P. Salati and R. Taillet, with whom we performed the 
very early studies on the topic. We are also indebted to L. Athanassoula, 
X.-J. Bi, F.-S. Ling, D. Maurin, E. Nezri, R. Teyssier and Q. Yuan for further 
more detailed and complete analyses achieved afterward. We finally 
thank the organizers of the conference for having offered such a nice 
environment, which warmly benefited to the interesting discussions and 
even debates hold in the sessions.



\end{document}